\documentclass[a4paper]{jpconf}
\usepackage{graphicx}
\usepackage{mathrsfs}
\begin{document}
\title{Measurement of the Inclusive and Fiducial Cross-Section of Single Top-Quark $t$-Channel Events
       in $pp$ Collisions at $\sqrt{s}$ = 8 TeV.}

\author{Phillipp Tepel On behalf of the ATLAS Collaboration.}

\address{Bergische Universit\"at Wuppertal, Fachbereich C Physik, Gau\ss strasse 21, 42119 Wuppertal, Germany}

\ead{tepel@physik.uni-wuppertal.de}

\begin{abstract}
This article presents the measurement of a $t$-channel single top-quark production
fiducial cross-section in the lepton+jets channel with 20.3~fb$^{-1}$ of 8~TeV data using a
neural-network discriminant.

A fiducial cross-section quoted within the detector acceptance of
$\sigma_{\rm fid} =  3.37 \pm 0.05 \, (\mathrm{stat.}) \pm 0.47 \, (\mathrm{syst.}) \pm 0.09 \, (\mathrm{lumi.})~$pb
is obtained.
The total inclusive $t$-channel cross-section is calculated using the acceptance
predicted by various Monte Carlo generators. If the acceptance from the
a{\sc MC@NLO} + \textsc{Herwig} event generator is used, a value of 
$\sigma_t = 82.6  \pm 1.2 \, (\mathrm{stat.})  \pm 11.4 \, (\mathrm{syst.}) \pm 3.1 \, (\mathrm{PDF}) \pm 2.3 \,
(\mathrm{lumi.})~$pb is obtained,
consistent with a theory calculation.  
Using the ratio of the measured inclusive cross-section to the predicted cross-section 
and assuming that the top-quark-related CKM matrix elements obey the relation
$|V_{tb}|\gg |V_{ts}|, |V_{td}|$, the coupling strength
at the $W$-$t$-$b$ vertex is determined to be
$|V_{tb}|=0.97^{+0.09}_{-0.10}$. Assuming that $|V_{tb}|\leq 1$
a lower limit of $|V_{tb}|>0.78$ is obtained at the 95\% confidence level.
\end{abstract}

\section{Introduction}
At the LHC the most dominant production of top quarks is given by the strong interaction. Beside the strongly produced $t\bar{t}$ pairs, singly produced top quarks emerge via the weak interaction.
Three production channels, of singly produced top quarks, are distinguished by the virtuality of the exchanged $W$ boson. The $t$-channel exchange of a virtual $W$ boson provides the leading contribution to the production rate at the LHC. 

This article presents the first measurement of the production cross-section in the fiducial volume in single top-quark topologies.
The measurement of the fiducial cross-section has advantages over a fully inclusive measurement. The fiducial cross-section has reduced model dependence, enabling direct comparisons between predictions, at the present and in the future. The reduced model dependence is also reflected in a lower total systematic uncertainty, compared to a fully inclusive cross-section.
Events are characterized by one isolated high-$p_{\mathrm{T}}$ charged lepton, exactly two jets one of which is required to be $b$-tagged and missing transverse momentum.
The main background processes are top quark associated diagrams, $t\bar{t}$ pair production as well as other single top-quark production channels. Processes with $W/Z$+jets, especially produced with heavy quark jets, diboson production and multijet production complete the background contributions.

\section{Event Selection}
Based on the expected signature of the single top-quark signal events, events are selected with exactly two jets, out of which one is identified as a $b$-quark jet, one isolated electron or muon and missing transverse momentum.

Offline electron and muon candidates are required to be isolated and satisfy $p_{\mathrm{T}}>25$~GeV and $|\eta|<2.5$.

Jets are clustered using the anti-$k_t$ algorithm with a cone parameter of $0.4$. Jets must fulfill $p_{\mathrm{T}}>30$~GeV and $|\eta|<4.5$ requirements. In the endcap-forward calorimeter region, $2.75<|\eta|<3.5$,jets are required to have $p_{\mathrm{T}}>35$~GeV. One of the selected jets is required to be identified as a $b$-quark jet by the MV1c $b$-tagging algorithm.

A set of cuts is placed to reduce the multijet background contribution. The transverse $W$-boson mass ($m_{\mathrm{T}}(W)$) must be higher than $50$~GeV as well as $E^{\mathrm{miss}}_{\mathrm{T}}>30$~GeV. A second set of cuts is imposed in the plane spanned by the  $p_{\mathrm{T}}$ of the lepton and the $\Delta\phi(j_1,\ell)$, see Equation \ref{eq:qcdcut}. 

\begin{equation}
p_{\mathrm{T}}>40\mathrm{GeV}\cdot \left(1-\frac{\pi-|\Delta\phi(j_1,\ell)|}{\pi-1} \right),
\label{eq:qcdcut}
\end{equation}
where $\ell$ denotes the selected lepton and $j_1$ is the leading jet in $p_{\mathrm{T}}$.

\section{Background Estimation}
The normalisation of the background processes is obtained from two different sources. For all processes, except the multijet contribution, the number of selected events is predicted using theory calculation. The normalisation of the multijet process is not predicted, and has to be obtained using a binned maximum likelihood fit to collision data. The model used for the multijet process is different for the electron and muon channel. In the electron channel, the jet-lepton method selects events containing a jet, which is likely to be misidentified as a lepton in the detector, from dijet monte carlo events. In the muon channel, the anti-muon method uses inverted identification criteria, to select fake lepton events from collision data. Both methods are explained in detail in \cite{ATLAS-CONF-2014-058}. The templates are used in the respective channels in a fit to the $E^{\mathrm{miss}}_{\mathrm{T}}$ distribution. For the fit, the requirement $E^{\mathrm{miss}}_{\mathrm{T}}>30$~GeV is omitted.

\section{Signal Discrimination}
To obtain a high separation between signal and background events, a neural network is employed. The neural network is trained with 14 observables which facilitate the most discrimination power between signal and background events. All 14 input variables have been tested for good agreement between data and simulation. Figure~(1) depicts the Neural Network output distribution, signal and backgrounds are normalised to the fit result of the Neural Network discriminant distribution.\\

\begin{minipage}{0.5\textwidth}
\includegraphics[width=16pc]{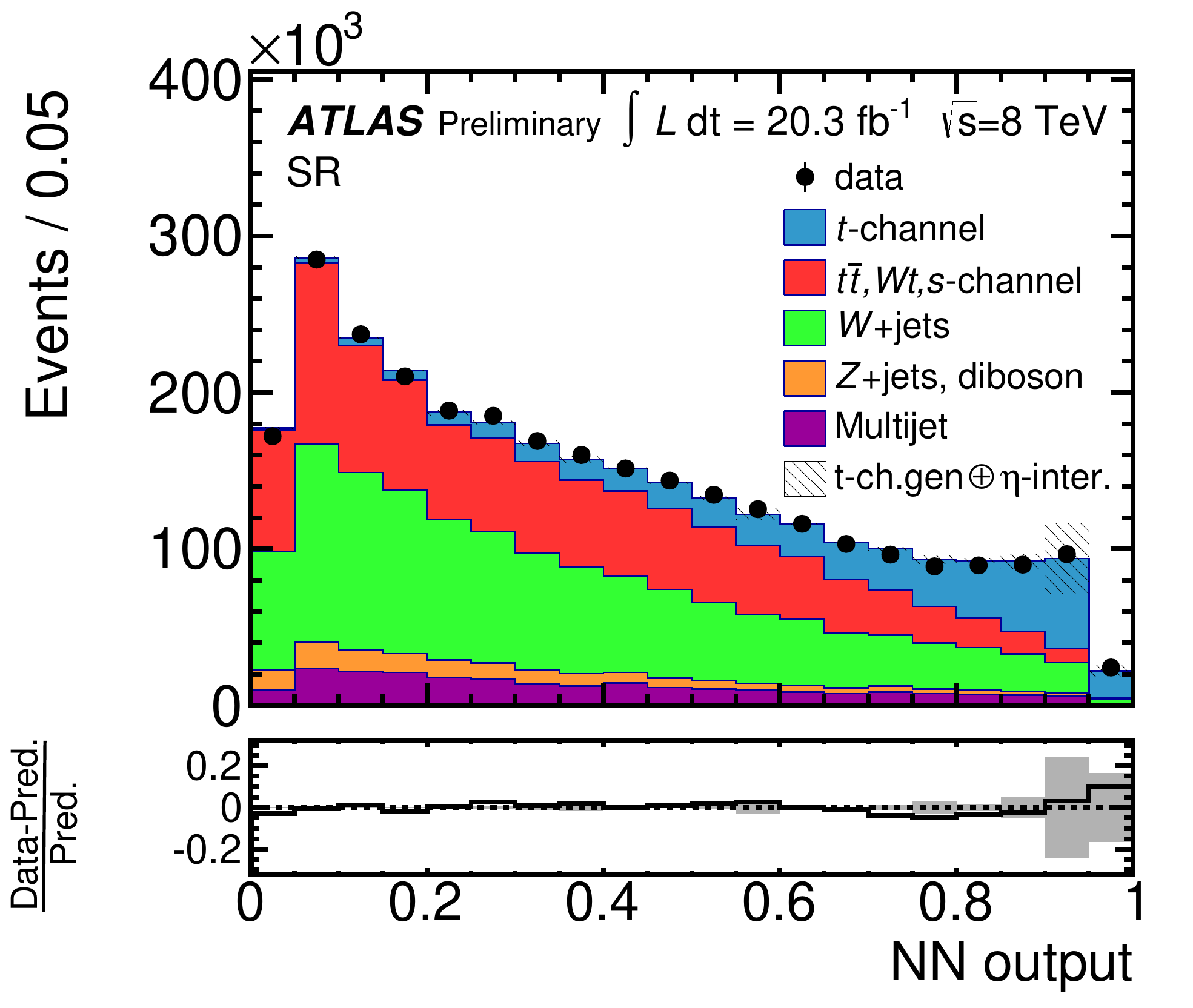}
\end{minipage}
\begin{minipage}{0.46\textwidth}
\textbf{Figure 1.} Neural-network output-distribution for the combined electron and muon channel. 
  The signal and background processes are normalised to the fit result of the Neural Network discriminant distribution.
   The bottom panel shows the relative difference between observed data and expectation. The grey shaded band 
   reflects the $t$-channel generator uncertainty and the uncertainty on the jet energy scale $\eta$-intercalibration. \cite{ATLAS-CONF-2014-007}
\end{minipage}

\section{Systematic uncertainties}
Systematic uncertainties on the normalisation of the individual
backgrounds and on the signal acceptance as well as uncertainties on the 
shape of the individual predictions affect the measured
$t$-channel cross-section. Both rate and
shape uncertainties are taken into account by generating correlated 
pseudo-experiments. The impact of the systematic uncertainties on the
$t$-channel cross-section measurement is estimated from these pseudo-experiments.
The classes of the considered uncertainties include objects modelling, monte carlo generator, PDF, background normalisation and Luminosity uncertainties.

\section{Cross-section Measurement}
The extraction of the $t$-channel single top-quark cross-section in this analysis 
is done in a fiducial volume within the detector acceptance. 
The fiducial cross-section $\sigma_{\rm fid}$ is defined as:
 \begin{eqnarray}
  \sigma_{\rm fid}=\frac{\epsilon_{\rm corr,sel}}{\epsilon_{\rm corr,fid}} \cdot \frac{\rm \hat{\nu}}{\cal{L}}=R_f\cdot \frac{\rm \hat{\nu}}{\cal{L}},
  \label{eq:acceptances}
\end{eqnarray}
where $\epsilon_{\rm corr,sel}$ is the fraction of events
which are selected by the offline selection to be within the fiducial volume and
$\epsilon_{\rm corr,fid}$ is the fraction of events within the fiducial volume to be selected
by the offline selection. 
The ratio of these two quantities is defined as $R_f$ and is of the order of 3.5.
The estimated number of expected $t$-channel single top-quark events is named
$\hat{\nu}$ and obtained by the binned maximum-likelihood fit.
The fiducial cross-section can then be extrapolated to the total inclusive cross-section $\sigma$ with:
\begin{eqnarray}
  \sigma=\frac{1}{\epsilon_{\rm fid}} \cdot \sigma_{\rm fid},
\end{eqnarray}
where $\epsilon_{\rm fid}$ is the selection efficiency of the particle-level selection.
The particle-level selection is imposed on objects defined on stable particles. The definition of the fiducial phase space is close to the phase space of the reconstructed and selected data set.

\section{Results}

The estimated number of $t$-channel single top-quark events $\hat{\nu}$ 
obtained from the binned maximum-likelihood fit of the network output distribution 
is used to calculate the fiducial cross-section.

\begin{eqnarray*}
\sigma_{\rm fid} &=&  3.37 \pm 0.05 \, (\mathrm{stat.}) \pm 0.47 \, (\mathrm{syst.}) \pm 0.09 \, (\mathrm{lumi.})~\mathrm{pb}.
\end{eqnarray*}
\\
The uncertainty on $\hat{\nu}$ is obtained from pseudo experiments and propagated
to the measured cross-section.  
The total relative uncertainty on the measurement of the fiducial cross-section is 
$\pm 14\%$, the dominant contributions being the uncertainty on the JES $\eta$-intercalibration and the $t$-channel
generator modelling.

A comparison of predictions for the fiducial cross-section for different NLO MC event generators 
and the matched LO generator \textsc{AcerMC} together with the measured value is shown in Figure~(2).
The fiducial cross-section is calculated using the inclusive cross section for each generator and the corresponding 
selection efficiency of the particle-level selection.

\begin{minipage}{0.5\textwidth}
\includegraphics[width=19pc]{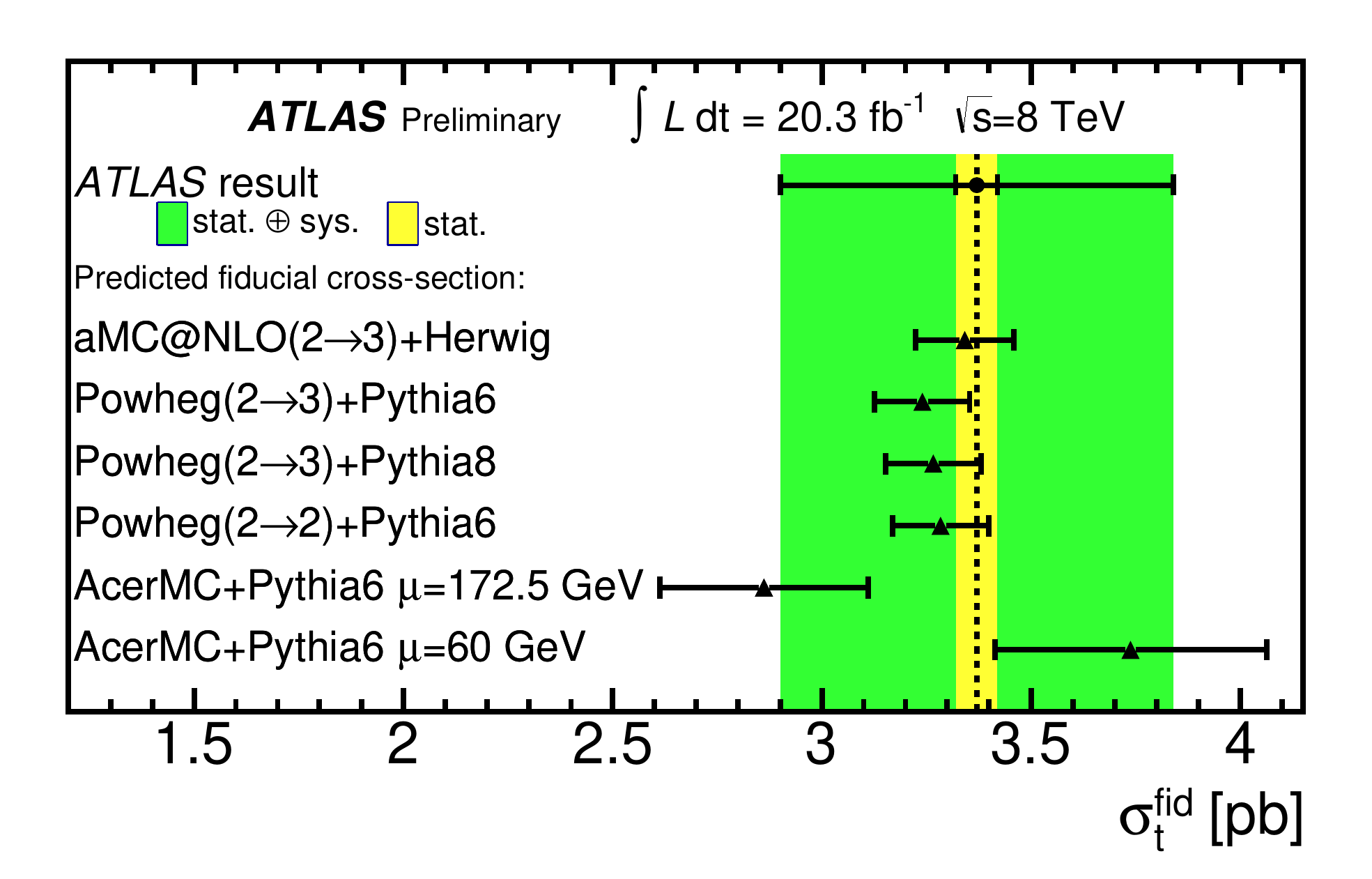}
\end{minipage}
\begin{minipage}{0.46\textwidth}
\textbf{Figure 2.} Predicted $t$-channel single top-quark fiducial cross-section for different NLO MC generators and the
matched LO generator \textsc{AcerMC} together with the measured value. The uncertainty on the prediction consists of the
scale uncertainty and the uncertainty on the PDFs. \cite{ATLAS-CONF-2014-007}
\end{minipage}

Using various MC generator models, the fiducial cross-section can be extrapolated 
to the full phase space and can be compared to the NLO+NNLL calculation. 
There is an additional uncertainty on the extrapolation due to the uncertainty on the PDF
with a size of 3.8 \%.
A summary of these inclusive cross-sections is presented in Figure~(3).
The values deduced with the acceptances obtained with the NLO generators
{\sc POWHEG} and a{\sc MC@NLO}  are in excellent agreement with the NLO+NNLL prediction 
and have only a difference of 1.7 \% to each other.
\vspace{20pt}

\begin{minipage}{0.5\textwidth}
\includegraphics[width=19pc]{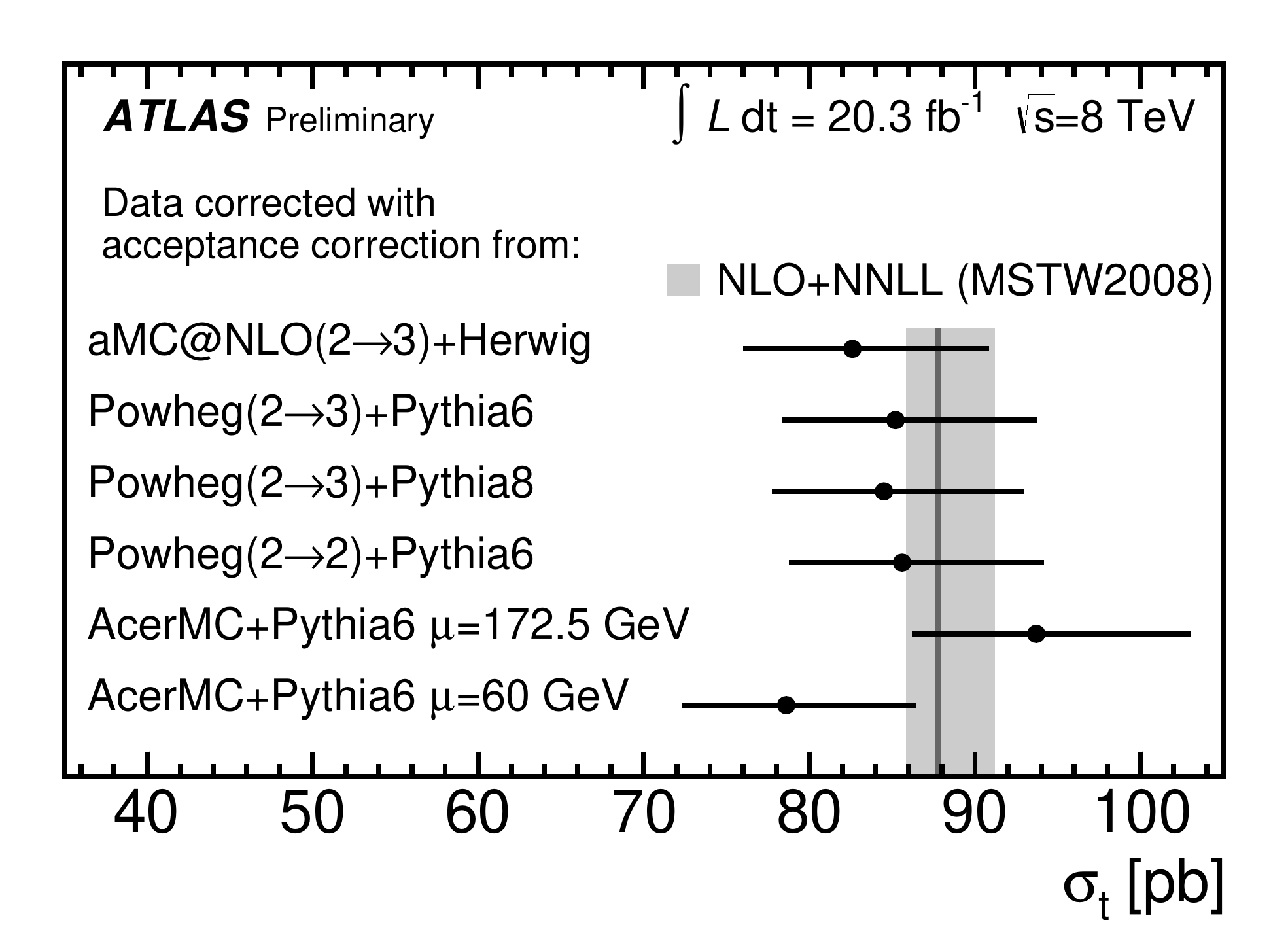}
\end{minipage}
\begin{minipage}{0.47\textwidth}
\textbf{Figure 3.} Inclusive total $t$-channel single top-quark production cross-section
         obtained from extrapolating the fiducial cross-section to the full phase space using the acceptance of different MC
         generators and taking the $\mathcal{B}(t\rightarrow \ell \nu b)$ into account.
         The vertical line indicates the NLO+NNLL theoretical cross-section, including the uncertainty displayed as a grey band.
         The uncertainty on each extrapolated cross-section emerges from the uncertainty on the fiducial cross-section and the uncertainty on the PDF. \cite{ATLAS-CONF-2014-007}
\end{minipage}
\\

Single top-quark production in the $t$-channel proceeds via a $W$-$t$-$b$ 
vertex and the measured cross-section is proportional to $|V_{tb}|^2$,
where $V_{tb}$ is the relevant CKM matrix element. 
The value of $|V_{tb}|$ is extracted by dividing the
extrapolated single top-quark $t$-channel cross-section using the acceptance of the a{\sc MC@NLO} + \textsc{Herwig} generators,
by the SM expectation~\cite{Kidonakis:2011wy}. 

\vspace{10pt}
The corresponding coupling at the $W$-$t$-$b$ vertex is determined to be 
$|V_{tb}|=0.97^{+0.09}_{-0.10}$ and the
95\% C.L. lower limit on the CKM matrix element $|V_{tb}|$ is 0.78.

\end{document}